\documentclass[12pt]{iopart}
\usepackage{graphicx}
\usepackage{amssymb}
\usepackage{bm}
\begin{document}
\title[Rate equations for epitaxial growth]
{Rigorous derivation of the rate equations for the epitaxial growth}
\author{V I Tokar$^{1,2}$ and H Dreyss\'e$^1$}
\address{$^1$IPCMS, UdS--CNRS, UMR 7504, 23 rue du Loess,
F-67034 Strasbourg, France}
\address{$^2$Institute of Magnetism, National Academy of Sciences,
36-b Vernadsky Boulevard, 03142 Kiev-142, Ukraine}
\ead{tokar@unistra.fr}
\begin{abstract}
In the framework of the second-quantization representation of the master
equation governing the irreversible epitaxial growth, exact equations
describing the evolution of the island densities has been obtained.
Their decoupling within a mean field-type approximation with the unknown
correlation functions replaced by capture numbers (CNs) has been used to
derive a closed set of rate equations.  The latter has been compared with
the exact equations to obtain rigorous definitions of the CNs. The CN that
describes the nucleation of dimer islands from two mobile monomers has
been measured in the exact kinetic Monte Carlo simulations with the use
of the rigorous definition. Strong disagreement with the literature values
calculated within alternative techniques has been found, especially at low
surface coverage. Plausible causes for the discrepancies are suggested.

Another important result of the rigorous approach is the rate
equation with the term describing the monomer diffusion.  The equation
significantly differs from the widely used equations of this kind known
from literature. Arguments in favour of our approach are given.
\end{abstract}
\pacs{68.55.A-,81.15.Aa}
\maketitle
\section{Introduction}
Lattice gas models (LGMs) with the gas atoms subject to stochastic
kinetics are widely used in modelling the coherent growth on the
surface \cite{barabasi,2006,jansen}.   The efficiency of the approach
comes from two sources. First, the coherent LGM configuration can be
fully characterized by the occupation numbers of a discrete lattice.
This greatly reduces the phase space of the system which in more refined
treatments is continuous and thus more difficult to deal with.  Second,
the stochastic kinetics of the LGMs allow for the use of the kinetic Monte
Carlo (KMC) technique which in the case of discrete variables is capable
of simulating the temporal evolution of the system on the experimental
time scale in contrast to more accurate molecular dynamics methods
which are restricted to very short times because of the computationally
demanding continuous phase \cite{2006,jansen}.

In practical surface growth studies, however, even the use of the KMC
meets with difficulties.  For example, one of the most frequently studied
quantities is the island size distribution \cite{2006} whose measurement
requires gathering statistics on the number of islands of all sizes.
Because in many cases the islands consist of several hundreds or even
thousands of atoms \cite{2006},  reliable statistics can be gathered
only in repeated simulations of large systems.  Therefore, sometimes
additional arguments need be invoked to interpret the simulated data
because of poor statistics \cite{albao_kinetic_2009}. Besides, there
exist situations when the physics of the problem requires simulation
of prohibitively large systems.  For example, in the cases when the
surface diffusivity is very large in comparison with the deposition rate,
the KMC simulations of systems smaller then some critical value would
lead to qualitatively incorrect results due to the finite size effects
\cite{1992,vvedensky_nano11}.  Accelerated algorithms were developed to
treat very large systems \cite{PRE08,PRB09} but their implementation in
concrete cases meets with difficulties \cite{amar10,vvedensky_nano11}.

Some of the difficulties with practical use of the KMC technique are
a direct consequence of its strength, that is, of its ability to give
a very detailed description of the system under study, such as, e.\
g., the precise morphologies of the growing islands.  Such descriptions
require considerable resources and efforts to obtain but are superfluous
in the cases when only a coarse-grained picture of the growth is needed.
For example, in \cite{brune_et_al} it was shown that to determine the
parameters characterizing the atomic diffusivity on the surface in
the sub-coalescence growth regime it is quite sufficient to gather
experimental information only on the island densities with their
morphologies being irrelevant.

Simplification of the modelling in such cases has been the
subject of extensive studies (see, e.\ g., review articles
\cite{ratsch-venables-review,2006}). It was found that the island
densities can be calculated with good accuracy in the framework of the
mean field-type rate equations (REs) whose accurate solution requires only
a fraction of the computational effort needed for the corresponding KMC
study (see \cite{venables1973,venables1984,1992,RE,2006,bales_chrzan,m_s}
and references therein).  But this efficiency comes at the price of
the necessity to independently assess many parameters or even functions
entering the REs.

The number and the nature of the parameters depend on the specifics of the
model under consideration.  In the irreversible growth model that will be
discussed in the present paper the input to the REs is a large (in theory
infinite) set of the so-called capture numbers (CNs) that describe the
rates with which islands of various sizes incorporate diffusing atoms (or
mobile monomers).  The CNs has been the subject of extensive investigation
since the introduction of the REs in this problem \cite{venables1984}.
In numerous studies a variety of functional forms and of the factors
influencing the CNs were researched, such as the power-law dependence
on the island size depending on the islands morphology \cite{RE}, on
the total island density (see \cite{1Dscaling,rzshvv} and references
therein), the influence of spatial correlations and capture zones,
\cite{approx_coincid,1996,2001,car_parking12}, of self-consistency
conditions and total coverage \cite{bales_chrzan,reversible}, etc..
Though considerable insights has been gained in these studies, fully
satisfactory solution has not yet be found so new ideas are being put
forward in in the search for accurate REs for the epitaxial growth
\cite{GWS,korner_island_2010,car_parking12}.

One of the difficulties hampering the development of the theory
is that in existing approaches the quality of the CNs is assessed
either indirectly via the quality of the island size distributions
they produce or with the use of an empirical procedure developed in
\cite{1996,2006,korner_island_2010}.   A major goal of the present
study is to propose a way to resolve this difficulty by suggesting
a method of a direct measurement of the CNs in the KMC simulations.
Our proposal is based on the observation that the stochastic KMC process
can be exactly described by the corresponding master equation (ME)
\cite{van_kampen,jansen}.   With the use of the ME we can derive exact
expressions describing the time evolution of the island densities.
These expressions can be transformed into a closed set of REs by
means of decoupling of a mean field kind.  The CNs that appear in the
decoupling can be compared to the exact expressions to obtain their
precise meaning.  With the use of these expressions the CNs will
be ``measured'' in the exact KMC simulations. In the course of the
derivation we will obtain an evolution equation containing the monomer
diffusion term also found in the self-consistent approaches to the CNs
calculation \cite{bales_chrzan,reversible,li_evans03,li_evans05}. Because
our equation differs from the known ones and arguably is more accurate,
we consider it as our second major achievement in this study. We hope that
in the future this equation will allow us to improve the self-consistent
theory of the CNs \cite{bales_chrzan,reversible,li_evans03,li_evans05}.

The paper is organized as follows. In the next section we introduce
the second quantization representation of the stochastic processes
that describe the coherent epitaxial growth; in section \ref{evolution}
the exact evolution equations are derived and the REs obtained via
their decoupling are presented in section \ref{rate_equations}; in
particular, in section \ref{REs} we show how the conventional
REs known from literature arise in our approach.
In section \ref{nucleation} we will measure the capture rate for the
nucleation of dimers in the so-called i=1 model of irreversible growth
\cite{2006}.  Because the capture rate for the creation of dimers
effectively defines the total island density \cite{bales_chrzan},
its accurate definition may improve the estimates of the surface
diffusion on the basis of experimental data \cite{brune_et_al}. In
the concluding section \ref{conclusion} we briefly discuss the
results obtained and possible farther development.
\section{Master equation in the second quantization representation}
The quantum formalism in application to stochastic LGMs  was introduced
in \cite{doi,siggia,grassberger_scheunert} and subsequently applied
by many authors to various stochastic models (for further bibliography
see \cite{mattis_glasser,PRE96}).  The formalism does not introduce any
new physics into the problem but allows one to efficiently deal with
huge matrices that appear in the ME approach in the kinetic many-body
problems using the powerful tools from the second quantization or the
quantum field theories.

For concreteness we will illustrate the formalism using the
simple model of submonolayer growth previously studied in, e.\ g.,
\cite{korner_island_2010}.  In this model the substrate has the square
lattice geometry.  It is assumed that each site can be occupied by one
atom at most.  Thus, the system consisting of $M$ sites the number of
configurations is $2^M$.
\subsection{The Fock representation of the state space}
Usually in LGMs the atomic configuration is characterized by the set
of the occupation numbers
\begin{equation}
\label{ occup_nos}
\{n_i=0,1\},
\end{equation}
where $i$ is the lattice site index.
But it is obvious that any quantity acquiring two values can be used.  In the
quantum formalism the state of the system under consideration which
we denote by the ``ket" vector $|\alpha\rangle$ can be characterized
by the product over all lattice sites
\begin{equation} \label{ vec}
|\alpha\rangle=\prod_i|b_i\rangle_i,\quad b_i=0,1
\end{equation}
of the binary vectors
\begin{equation} \label{ binary}
|0\rangle_i=\left(\begin{array}{c}0\\1\end{array}\right)_i
\mbox{\ \ and\ \ }
|1\rangle_i=\left(\begin{array}{c}1\\0\end{array}\right)_i,
\end{equation}
where the first vector corresponds to the empty site and the second
vector to the occupied one.  Now, by introducing corresponding conjugate
``bra"-vectors via the transposition of vectors (\ref{ binary}) one
can see that the set of state vectors (\ref{ vec}), besides being
complete, is also orthonormal.

Transitions between the states of different occupation in  (\ref{
binary}) are performed with the use of the creation and annihilation
operators of the hard-core bosons \cite{grassberger_scheunert,PRE96}
\begin{equation}
\label{ aa+}
a^+_i=
\left(\begin{array}{cc}
0 & 1\\
0 & 0
\end{array}\right)_i
\mbox{\ \ and\ \ }
a_i =
\left(\begin{array}{cc}
0 & 0\\
1 & 0
\end{array}\right)_i
\end{equation}
In the space of vectors (\ref{ vec}) of dimensionality $2^M$ the matrices
have the size $2^M\times2^M$.  Therefore, the action of operators (\ref{
aa+}) should be augmented with unit $2\times2$ matrices acting on the
sites different from $i$.  For simplicity in derivations below we omit
these trivial factors.

As is easily verified with the use of the explicit representations above,
$a_i$ and $a^+_i$ satisfy all the usual properties of the creation and
annihilation operators
\begin{equation}
\label{ avec}
a^+_i|0\rangle_i=|1\rangle_i,\qquad a_i|1\rangle_i=|0\rangle_i.
\end{equation}
In particular, the operator whose eigenvalues correspond to the
occupation numbers  (\ref{ occup_nos}) has the conventional form
(for simplicity here and below we will use the same notation for the
operator and for the occupation numbers)
\begin{equation}
\label{ n}
n_i=a^+_ia_i=
\left(\begin{array}{cc}
1 & 0\\
0 & 0
\end{array}\right)_i
\end{equation}
as is easily seen from  (\ref{ aa+}).
Despite the no-double-occupancy conditions that follow from the
definitions  (\ref{ aa+}) and are reminiscent of the fermions:
\begin{equation}
\label{ no2occupancy}
\left \{ a^+_i, a_i\right \}=1 \mbox{\ and\ } (a^+_i)^2=a_i^2=0
\end{equation}
(the braces denote the anticommutator), the operators at different sites
commute:
\begin{equation}
\label{ commutation}
\left [ a^+_i,a_j \right ]_{i\neq j}=0.
\end{equation}
Finally, by defining the vacuum state of the system as that of the empty
lattice
\begin{equation} \label{ vac}
|0\rangle = \prod_i|0\rangle_i
\end{equation}
we can represent the state vectors (\ref{ vec}) of the system as the
Fock space
\begin{equation} \label{ fock}
|\alpha\rangle = \left(\prod_{\mbox{\scriptsize Occupied sites}}a^+_i
\right)|0\rangle.
\end{equation}
\subsection{Master equation in the Fock space}
The stochastic LGM of the surface growth that we consider
in the present paper belongs to the general category of stochastic models
whose kinetics satisfy the ME
\begin{equation}
\label{ ME}
\frac{dP_\alpha}{dt}=\sum_\beta \left(W_{\alpha\beta}P_\beta
-W_{\beta\alpha}P_\alpha\right),
\end{equation}
where $P_\alpha(t)$ is the probability of the system to be found at
time $t$ in state $\alpha$ and $W_{\alpha\beta}$ is the matrix of
transition rates between different states of the system.  As was shown
above, in the case of the growth on the surface in the representation
of occupation vectors (\ref{ binary}) the matrix has $M$ binary indices
and its size is $2^M\times2^M$.  Dealing with such matrices in the case
of macroscopic surfaces [$M\sim O(10^{15})$] is not an easy task. But
as we saw on the example of the creation and annihilation operators
(\ref{ aa+}), the structure of the $2^M\times2^M$ matrix can be rather
trivial consisting mostly from the Kronecker product of the $2\times2$
identity matrices.  The Fock space formalism emphasizes the nontrivial
aspects of the transition matrix leaving trivialities behind the scene.
For example, a simple model of the surface growth consisting only in
atomic deposition can be described within this formalism by the operator
\begin{equation}
\label{ W_F}
\mathbf{W}^F = F\sum_{i}a^+_i,
\end{equation}
where $F$ is the deposition rate.  Using explicit representation
for the state vectors  (\ref{ vec}) and the operator identities
(\ref{ no2occupancy}) it is easy to see that the matrix element
\begin{equation}
\label{ W_F2}
{W}^F_{\alpha\beta}\equiv \langle\alpha|\mathbf{W}^F|\beta\rangle=
\left\{\begin{array}{l}0\\F\end{array}\right.
\end{equation}
is different from zero only provided $|\alpha\rangle=a^+_i|\beta\rangle$
for some site index $i$ and the site is empty in state $|\beta\rangle$
because according to  (\ref{ no2occupancy}) two atoms cannot sit on
the same site.  Thus, the transition matrix in this case is very sparse
despite being very large.  Similarly, hopping diffusion of a free atom
on the surface can be described with the use of the operator
\begin{equation}
\label{ W_D0}
\mathbf{W}^D_0 = D\sum_{ii_{\rm N}}a^+_{i_{\rm N}}a_i,
\end{equation}
where $D$ is the diffusion rate and we assumed that the atom that left
site $i$ may hop only to one of four nearest sites denoted here and in
the following by the subscript ${\rm N}$.

Further, using the complete set of the system states, it is
convenient to cast the ME (\ref{ ME}) into the form
\begin{equation}\label{ ME2}
\frac{d|t\rangle}{dt}=\mathbf{T}|t\rangle,
\end{equation}
where
\begin{equation}
\label{ t}
|t\rangle =\sum_{\alpha}P_\alpha(t) |\alpha\rangle
\end{equation}
and
\begin{equation}
\label{ T}
\mathbf{T}=\left[W_{\alpha\beta}-\delta_{\alpha\beta}
\sum_\gamma W_{\gamma\alpha}\right].
\end{equation}
In practice one is usually interested in the average values of various
densities described by the operators that are diagonal in the Fock space
and can be calculated as
\begin{equation}\label{ Omean}
O(t)=\langle \hat{O}\rangle(t)=\sum_\alpha P_\alpha(t)
\langle\alpha| \hat{O}|\alpha\rangle.
\end{equation}
To express such averages in the form of matrix elements, as is conventional
in quantum formalism, we introduce the matrix containing all possible states
of the system with equal (unit) weight as
\begin{equation}
\label{ Omega}
|\Omega\rangle = \sum_{\beta} |\beta\rangle.
\end{equation}
Now using the orthonormality of the state vectors one gets
\begin{equation}
\label{ Omean2}
O(t)=\langle\Omega|\hat{O}|t\rangle.
\end{equation}
Another important observation concerning vector $|\Omega\rangle$ in
(\ref{ Omega}) is that, as can be seen from  (\ref{ T}), the sum of
the matrix elements of matrix $\mathbf{T}$ over the left index is equal
to zero: $\sum_{\alpha}{T}_{\alpha\beta}=0$.  This is the consequence
of the fundamental requirement of the probability conservation and in
the quantum notation can be written as
\begin{equation}
\label{ conservation}
\langle\Omega|\mathbf{T}=0.
\end{equation}
Now using this identity and formally solving  (\ref{ ME2}) via the
operator exponent and assuming that the initial state at $t=0$ is the
vacuum state  (\ref{ vac}) (the empty lattice) one can re-write
(\ref{ Omean2}) as
\begin{equation}
\label{ Omean3}
O(t)=\langle\Omega|e^{-t\mathbf{T}}\hat{O}e^{t\mathbf{T}}|0\rangle.
\end{equation}
From this we finally obtain an exact equation for the time evolution
of the average value of $O(t)$:
\begin{equation} \label{ exact}
\frac{dO}{dt}=\langle\Omega|[\hat{O},\mathbf{T}]|t\rangle.
\end{equation}
\section{\label{evolution}Evolution of island densities during submonolayer 
growth}
To derive the evolution equations for the island densities one needs to
know the operators $\mathbf{T}$ and $\hat{O}$ entering  (\ref{ exact}).
We first note that it suffices to know only the first non-diagonal
term of $\mathbf{T}$ in  (\ref{ T}) because the density operators we
are interested in are diagonal and thus will commute with the diagonal
part in  (\ref{ exact}).  (In case of necessity the diagonal part of
$\mathbf{T}$ can be easily recovered from the non-diagonal part given
below \cite{PRE96}.) The operator $\mathbf{W}$ that describes the
epitaxial growth in the model under consideration consists of only two
terms: the deposition term  (\ref{ W_F}) and the diffusion term of the
type of (\ref{ W_D0}) modified to include interatomic interactions through
the dependence of the diffusion rate on the atom environment:
\begin{equation}
\label{ totalW}
\mathbf{W}=F\sum_{i}a^+_i+\sum_{i,i_{\rm N}} a^+_{i_{\rm N}}D_i{a}_i.
\end{equation}
This is done with the use of the activated dynamics that is conventionally
used in the surface growth
studies \cite{clarke:2272,bales_chrzan,ratsch-venables-review}:
\begin{eqnarray}
\label{ D_i}
&&D_i=D\exp[-(E_N/kT)\sum_{i_{\rm N}}n_{i_{\rm N}}]
=D\prod_{i_{\rm N}}\exp[-(E_N/kT)n_{i_{\rm N}}]\nonumber\\
&&=D\prod_{i_{\rm N}}\left(1+\left[\exp(-E_N/kT)-1\right]
n_{i_{\rm N}}\right)\stackrel{E_N\gg kT}{\longrightarrow}
D\prod_{i_{\rm N}}\bar{n}_{i_{\rm N}}.
\end{eqnarray}
Here $E_N>0$ is the energy of binding between neighbour atoms, in
the second row use has been made of the property $n_i^2=n_i$, and
$\bar{n}_{i}=1-{n}_{i}$.  The irreversible growth corresponds to the
limit shown at the end of  (\ref{ D_i}) so in the calculations below
we will use the diffusion term in the form
\begin{equation}
\label{ W_D}
\mathbf{W}^D = D\sum_{ii_{\rm N}}a^+_{i_{\rm N}}\tilde{a}_i,
\end{equation}
where according to  (\ref{ D_i})
\begin{equation}
\tilde{a}=a_i\prod_{i_{\rm N}}\bar{n}_{i_{\rm N}}.
\end{equation}

\begin{figure}
\begin{center}
\includegraphics[viewport = -20 0 195 148, scale=1.25]{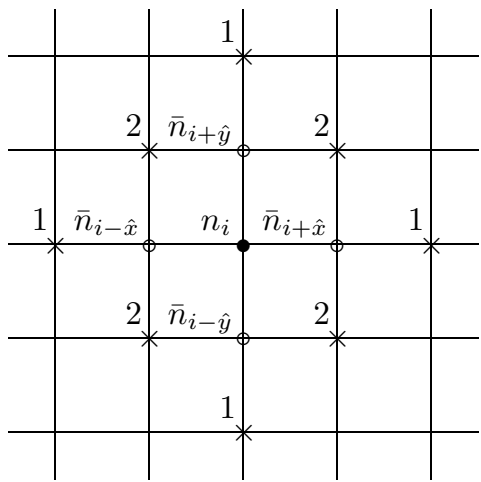}
\end{center}
\caption{Encircled sites illustrate the structure of the monomer operator
$\hat{N}_{1,i}$ in \eref{ N_1i}.  The crossed sites are the next
neighbours (NN) to site $i$ and the numbers are the values of $m_{i_{\rm
NN}}$ (the numbers of nucleation paths) in \eref{ monomers_t}.}\label{fig1}
\end{figure}

Next we need to define the island density operators.  We define atomic
islands on the square lattice similar to the definition of the lattice
animals \cite{jensen_enumerations_2001},  namely, as the clusters of
atoms occupying the sites any pair of which can be connected by a path
that traverses only nearest neighbour sites all belonging to the cluster.
We characterize the island by its size $s$, configuration $c$, and the
position on the lattice $i$.  The latter can be defined arbitrarily
but uniquely for all island of given configuration.  The islands have
different configurations if they cannot be obtained from each other by
lattice translations.  According to this definition, the dimers oriented
along axis $x$ and along axis $y$ have different configurations.  Thus,
the density operator for the cluster of any size (including both the
islands and the monomers) that has configuration $c$ and is placed at
cite $i$ is
\begin{equation}
\label{ islandN}
\hat{N}^{(c)}_{s,i}=\prod_{\{j\in c\}}n_{j}
\prod_{\{j_{\rm N}\in\bar{c}\}}\bar{n}_{j_{\rm N}}.
\end{equation}
The second product in this expression is over the sites bordering the
island by which we mean all sites that do not belong to the island but
are nearest neighbour to at least one island site (we will denote the
set of such sites as $\bar{c}$).  Operators $\bar{n}_{j_N}$ are needed to
guarantee that the island contains exactly $s$ atoms: without this factor
the operator $\hat{N}^{(c)}_{s,i}$ could produce nonzero expectation
value at configurations $c$ that make parts of larger islands. In the
case of monomer operator $\hat{N}_{1,i}$ the superscript $c$ is not
needed because the monomer configuration is unique (see \fref{fig1}):  
\begin{equation} \label{ N_1i}
\hat{N}_{1,i}={n}_i\prod_{\hat{e}=\pm\hat{x},\pm\hat{y}}\bar{n}_{i+\hat{e}},
\end{equation} 
where $\hat{x}$ and $\hat{y}$ are the unit lattice vectors.
\subsection{Evolution equations for the island densities}
Substituting the island density operators  (\ref{ islandN}) into (\ref{
exact}) and calculating the commutator (see \ref{commutators}) one arrives
on the evolution equations for their average values.  The densities are
calculated according to the usual expressions:
\begin{eqnarray}
\label{ meanNc}
&&N^{(c)}_s=M^{-1}\sum_{i}N^{(c)}_{s,i},\\
&&\langle\hat{N}_{1,i_{\rm NN}}\hat{N}^{(c)}_{s}\rangle=
M^{-1}\sum_{i}\langle\hat{N}_{1,i_{\rm NN}}\hat{N}^{(c)}_{s,i}\rangle,
\,\mbox{\em \ etc.}
\label{ meanN1Nc}
\end{eqnarray}
In the first equation the carets over the symbols are omitted because
here they are not operators but the number densities calculated according
to (\ref{ Omean2}); in the second equation and in the following the
subscripts $i_{\rm N}$ or $i_{\rm NN}$ are considered to be relative
coordinates with respect to the corresponding cluster or site (nearest-
or next-nearest neighbours) and thus in  (\ref{ meanN1Nc}) they are not
averaged out.

Thus, for the islands ($s>1$)  (\ref{ exact}) with
the use of (\ref{ Dcommutator}) and (\ref{ Fcommutator}) takes the
form
\begin{equation}
\fl\label{ islands_t}
\frac{dN^{(c)}_s}{dt}=
D\!\!\!\!\sum_{\{c\setminus 1\}i_{\rm NN}}^{\qquad\prime}
\langle\hat{N}_{1,{i_{\rm NN}}}\hat{N}^{(c\setminus 1)}_{s-1}\rangle
-D\!\sum_{{i_{\rm NN}}}m_{i_{\rm NN}}\langle
\hat{N}_{1,{i_{\rm NN}}}\hat{N}^{(c)}_{s}\rangle
+F\!\sum_{\{c\setminus1\}}{N}^{(c\setminus1)}_{s-1}
-F\bar{s}_c{N}^{(c)}_{s}+\dots.
\end{equation}
The prime over the first sum reminds that the summation over $i_{\rm
NN}$ is not over all NN sites to the $s-1$-atom island but only over
the NN positions that the detached atom can reach and remain free.
By the dots we denoted the terms responsible for the coalescence.
It would not be too difficult to write down these terms.  Because of
the symmetry of the square substrate a diffusing atom can be caught by
one, two or three nearby clusters (islands and/or monomers).  Similarly,
the process of direct impingement can include up to four clusters.  Thus,
as can be seen, these processes could have been accounted for in \eref{
islands_t} with the use of 19 additional terms containing all admissible
combinations of islands and monomers.  Inclusion of these terms, however,
would have made the equation very cumbersome.  Besides, currently the
research activity in the field of the irreversible growth is centred
around the pre-coalescence regime where only the processes taken into
account in \eref{ islands_t} participate.  This regime corresponds to
smaller coverages $\theta \lesssim 0.2$ and physically describes the
ensemble of well separated isolated islands interspersed with low density
monomer gas.  Because the density of all clusters are very low the processes
between closely spaced three or four clusters are small and can be neglected
in this regime.  If, however, there is need to account for such processes
(at high coverages, for example), our technique allows for rigorous derivation
of necessary equations.  But in the present paper we will neglect
the coalescence terms in all equations below.

In order to include spatial correlations and/or fluctuations into the
mean-field description, equation describing the monomer diffusion have
been proposed \cite{bales_chrzan,reversible,li_evans05,vvedensky90diff}.
To obtain similar equation in our approach we postpone the site averaging
in \eref{ meanNc} corresponding to $\hat{N}_{1,i}$ till later. Now from
(\ref{ Dcommutator1}) and (\ref{ Fcommutator1}) one gets
\begin{equation}
\label{ monomers_t}
\fl\frac{dN_{1,i}}{dt}=F{N}_{0,i} -4F{N}_{1,i}
+D\left({\sum_{i_{\rm N}}}^{\prime}{N}_{1,i_{\rm N}}
-4{N}_{1,i}\right)-\sum_{i_{\rm NN}}m_{i_{\rm NN}}\langle
\hat{N}_{1,{i_{\rm NN}}}\hat{N}_{1,i}\rangle,
\end{equation}
where ${N}_{0,i}$ is defined in \eref{ empty} and the prime on the
summation over $i_{\rm N}$ means that the corresponding terms are
present only if site $i$ does not have nearest neighbour atoms.  This is
a consequence of the irreversible dynamics we are studying.  If the atom
at $i$ had a neighbour it would have been bound and so could not produce
a free monomer via the diffusion hop.

The comparison of \eref{ monomers_t} with the equations proposed in
\cite{bales_chrzan,li_evans05} as well as derivation with its help of
the conventional site averaged rate equation for $N_1$ will be given in
the next section.
\section{\label{rate_equations}Rate equations}
The equations (\ref{ islands_t}) and (\ref{ monomers_t}) cannot be used
directly for the calculation of island densities because they contain
unknown terms---the averages of the operator products in the angular
brackets.  The standard way to proceed in such cases is to decouple
them by replacing the mean of the operator product with the product of
the means
\begin{equation}
\label{ MF}
\langle\hat{N}_{1,i_{\rm NN}}\hat{N}^{(c)}_{s}\rangle\approx N_1{N}^{(c)}_{s}.
\end{equation}
This is the conventional mean field approximation that proved its
efficiency in a lot of many-body problems.  In the
theory of epitaxial growth, however, the standard mean-field approach was
found to be too crude \cite{2006}.  Therefore, it was suggested to account
for the correlations neglected in  (\ref{ MF}) via the use of
parameters $\sigma^{(c)}_{s}$ defined according to the equality
\begin{equation}
\label{ sigma}
\sum_{i_{\rm NN}}\langle\hat{N}_{1,i_{\rm NN}}\hat{N}^{(c)}_{s}\rangle=
\sigma^{(c)}_{s}N_1{N}^{(c)}_{s}.
\end{equation}
Because of their physical meaning the parameters are called the capture
numbers \cite{2006}.  Substituting  (\ref{ sigma}) into (\ref{ islands_t})
and augmenting the set by the equation for the monomer density derived
in \ref{REs} we arrive at a formally closed set of equations that
describe the evolution of the island densities.  The set, however,
is not complete without explicit expressions for the capture numbers.
As we noted in the Introduction, there exist many heuristic approaches
aiming to define the CNs independently.  Our aim in the present paper
is to propose a method to derive the exact values of the CNs from
the KMC simulation data.  This would provide us with a definite mean
of assessing the quality of the CNs obtained in empirical approaches.
We achieve this by noting that equations (\ref{ islands_t}) and (\ref{
monomers_t}) are rigorously derived from the ME  (\ref{ ME}) that exactly
describes the KMC stochastic process.  Thus, the KMC data should satisfy
these equations exactly hence by comparing them with the empirical REs
one can calculate the exact values of CNs.  For example,  (\ref{ sigma})
can be trivially solved as
\begin{equation} \label{ sigma_c}
\sigma^{(c)}_{s}=\sum_{i_{\rm NN}}\langle\hat{N}_{1,i_{\rm NN}}
\hat{N}^{(c)}_{s}\rangle/N_1{N}^{(c)}_{s}.
\end{equation}
We note that all quantities on the right hand side of this expression
can be measured in the course of exact KMC simulations thus providing
the exact values of CNs.  A comment is in order concerning the influence
on this result of the coalescence terms that we omitted in our evolution
equations.  It should be noted that they also are discarded in the REs
that we are studying in this paper \cite{1992,2006} and that the CNs
cannot be calculated with better accuracy than their definition allows.
Thus, for the equations derived in the ``no coalescence'' approximation in
the KMC simulations we may count only those configurations where exactly
one monomer and one island participate and discard those processes which
lead to coalescence.  In this sense the measurements of CNs with the use
of \eref{ sigma_c} and KMC may be considered to be in exact correspondence
with its definition.

Individual capture numbers for every island configuration introduced
in (\ref{ sigma}) and (\ref{ sigma_c}) can be efficient in the cases
when the configurations are easily enumerable,  e.\ g., in the case of
one-dimensional islands either on the one-dimensional (1D) substrate
or on 2D substrates in the cases of very anisotropic quasi-1D growth
\cite{nat_chains,nat_chains2,albao_kinetic_2009}.  Because 1D islands have
only one configuration characterized by their size $s$, the superscript
$(c)$ for such islands in the above equations is superfluous.

In more common case of isotropic growth in 2D, however, the number
of island configurations is so large that for $s\gtrsim50$ even their
enumeration is infeasible \cite{jensen_enumerations_2001}.   In such cases
cruder description of island morphologies is needed.  Usually only the
island size statistic is being gathered both in the KMC simulations and
in the growth experiments.  So now we derive the formulas generalizing
(\ref{ sigma_c}) to this case.
\subsection{\label{REs}REs in isotropic case} 
Because the number of configurations of islands of sizes $s\gtrsim50$
is unmanageable \cite{jensen_enumerations_2001}, the rate equations are
usually written for the total density of islands containing the same
number of atoms as
\begin{equation}
\label{ meanNs}
N_s=\sum_{c}N^{(c)}_s.
\end{equation}
The standard REs for the island densities during the irreversible
growth have the form \cite{2006}
\begin{equation}
\label{ res}
\fl\frac{dN_s}{dt}=DN_1(\sigma_{s-1}N_{s-1}-\sigma_{s}N_{s})
+F\kappa_{s-1}N_{s-1}-F\kappa_sN_s,\qquad s=2,3,\dots
\end{equation}
We can derive them from (\ref{ islands_t}) by summing the latter over
the configurations corresponding to the same island size as in  (\ref{
meanNs}).  In this way we obtain the left hand side of  (\ref{ res}).
But on the right hand side one finds the terms of the kind
\begin{equation}
\label{ kappa-}
-F\sum_c\bar{s}_c{N}^{(c)}_{s}
\end{equation}
[see the last term in  (\ref{ islands_t})]
Identifying it with the last term in  (\ref{ res}) we arrive at the
following definition of the capture numbers for direct impingement
\begin{equation}
\label{ kappas}
\kappa_s=\sum_c\bar{s}_c{N}^{(c)}_{s}/{N}_{s}.
\end{equation}
One may wonder whether this definition is consistent with the positive
deposition terms in (\ref{ islands_t}) and (\ref{ res}) because the
positive term in  (\ref{ islands_t}) has the form quite different from
that of the negative term.  To see that they acquire the same form after
the summation over configurations it is sufficient to note that every term
of the kind ${N}^{(c\setminus1)}_{s-1}$ will enter the sum proportional
to $F$ in (\ref{ islands_t}) as many times as there is configurations
with $s$ atoms that would lead to the configuration $(c\setminus1)$
when the atom is taken away.  Obviously that the atoms can be taken only
from one of the sites in the configuration border so there is exactly
$\overline{(s-1)}_{(c\setminus1)}$ such terms.

Similarly, using the second term on the first line of  (\ref{ islands_t})
and comparing it with \eref{ res} one gets
\begin{equation}
\label{ sigmas}
\sigma_s=\sum_{i_{\rm NN,c}}m_{i_{\rm NN}}
\langle\hat{N}_{1,{i_{\rm NN}}}\hat{N}^{(c)}_{s}/N_1N_s.
\end{equation}
The contribution of the first (positive) term in  (\ref{ islands_t})
can be analysed similar to the direct impingement case discussed above.

To close the set of the rate equations for the island densities, the
equation for $N_1=M^{-1}\sum_{i}N_{1,i}$ is needed.  Its left hand
side is trivially obtained by averaging the left hand side of \eref{
monomers_t} over the surface, so let us consider in detail the right hand
side of this equation.  The first term the ``empty island'' operator
gives a nonzero contribution equal to unity only when site $i$ and all
its neighbours are empty.  This excludes all occupied sites as well as
the sites bordering the islands and the monomers.  Taking into account
(\ref{ kappas}) and the second term in  (\ref{ monomers_t}) which is
simply an additional monomer contribution $-\kappa_1N_1$ ($\kappa_1=4$)
we obtain the term in the parentheses of the conventional RE for the
monomers (see, e.\ g., \cite{korner_island_2010}):
\begin{equation}
\fl\label{ re1}
\frac{dN_1}{dt}=F\left(1-\theta-2\kappa_1N_1
-\sum_{s>1}\kappa_sN_s\right)
-2D\sigma_1N_1^2 -DN_1\sum_{s>1}\sigma_sN_s.
\end{equation}
The diffusion term in  (\ref{ monomers_t}) is the discrete Laplacian that
consists of two terms: the positive contribution from the monomer hops
on site $i$ from four neighbour sites with monomer densities $N_{1,i_{\rm
N}}$ and the negative contribution from site $i$ on the neighbour sites.
Thus, when summed over the whole lattice the negative terms tend to be
cancelled by the positive contributions from the neighbour sites.  In the
presence of islands and other monomers, however, the positive terms may
not exist because of the occupied sites nearest to them.  This leaves
uncompensated terms in the negative contribution that equals exactly
to the number $m_{i_{\rm NN}}$ of the nucleation paths to the mentioned
islands or monomers \cite{reversible}.  This in accordance with  (\ref{
sigmas}) gives the last two negative terms in (\ref{ re1}).  Finally,
the contribution that gives factor 2 to $\sigma_1$ term comes from the
last term of  (\ref{ monomers_t}).
\subsection{Comparison with equations from the 
self-consistent approach to the CNs calculation \cite{bales_chrzan}}
To facilitate the comparison we first rewrite \eref{ monomers_t} in the  
continuous substrate coordinates $i\to\mathbf{x}$ as
\begin{equation} \label{ free_mnmr}
\frac{d N_1\mathbf{(x)}}{dt} \simeq FN_0\mathbf{(x)} -\kappa_1FN_1\mathbf{(x)} + D\nabla^2N_1\mathbf{(x)}
-D\sigma_1N_1\mathbf{(x)}^2,
\end{equation}
where $N_0\mathbf{(x)}$ is the continuous version of the ``empty island''
operator $\hat{N}_{0,i}$ from \eref{ monomers_t}.

In this notation equation (11) from \cite{bales_chrzan} would read (we
rearranged some terms for convenience)
\begin{eqnarray}\label{11}
\frac{d N_1\mathbf{(x)}}{dt} \simeq&& F -\kappa_1FN_1\mathbf{(x)} 
+ D\nabla^2N_1\mathbf{(x)} -2D\sigma_1N_1N_1\mathbf{(x)}\nonumber\\
&&-F\sum_{s=1}^{\infty}\kappa_sN_s-DN_1\mathbf{(x)}\sum_{s>1}^{\infty}
\sigma_sN_s,
\end{eqnarray}
where
\begin{equation} \label{avrgN1}
N_1\equiv\frac{1}{Ma^2}\int d\mathbf{x}N_1\mathbf{(x)}
\simeq\frac{1}{M}\sum_iN_{1,i}
\end{equation}
(the lattice constant $a$ is assumed to be equal to unity). In comparing
\eref{11} with \eref{ free_mnmr} we first note that the last term
in \eref{ free_mnmr} that describes the island nucleation is local
in $\mathbf{x}$. This is reasonable because the adatoms nucleate an
island when meeting at one spatial point in the continuum approximation
(or at the nearest sites on the lattice) while according to \eref{11}
and \eref{avrgN1} the adatom at point $\mathbf{x}$ nucleates an island
with any other adatom in the system which is unphysical.  Secondly, in
\eref{11} the nucleation term enters with the coefficient 2, in contrast
to our equation \eref{ free_mnmr}.  The origin of this discrepancy
is connected with the last term of \eref{11} that is absent in our
equation \eref{ free_mnmr} but is present as the last term in  \eref{
re1}).  As we saw in section \ref{REs}, it arises from the averaging the
discrete Laplacian in \eref{ monomers_t} over the whole surface. Thus,
in the continuum approximation this term also should appear from the
integration of $\nabla^2N_1\mathbf{(x)}\equiv \mbox{div}\left[\nabla
N_1\mathbf{(x)}\right]$ as the sink term for the monomer attachment to the
island boundaries.  But as it is already present in  \eref{11} this means
that the capture of monomers is counted twice in this equation, unless
some very special treatment of the boundary conditions is undertaken
which seems to be the case in \cite{bales_chrzan}.

Similarly, the first term on the second line in \eref{11}
appears in our formalism only after the first term in \eref{
free_mnmr} is averaged over the whole surface, as in \eref{ re1}).
These direct impingement terms are less important under the usual
growth conditions and are often neglected, though they may became
important in some cases.  Apart from these terms, our equation \eref{
free_mnmr} is very similar to the $i=1$ case of the equation derived in
\cite{li_evans03,li_evans05}. In this approach the nucleation is local
and the capture term $DN_1\mathbf{(x)}\sum_{s>1}^{\infty} \sigma_sN_s$
is absent from equation (2.7) in \cite{li_evans05}.  The nucleation term,
however, enters with the coefficient $i+1=2$ which we consider to be
double counting. Careful comparison of all three equations with exact
KMC simulations is necessary to clarify this complicated issue.
\section{\label{nucleation}The nucleation rate}
\begin{figure}
\begin{center}
\includegraphics[viewport = 110 0 300 280, scale=0.55]{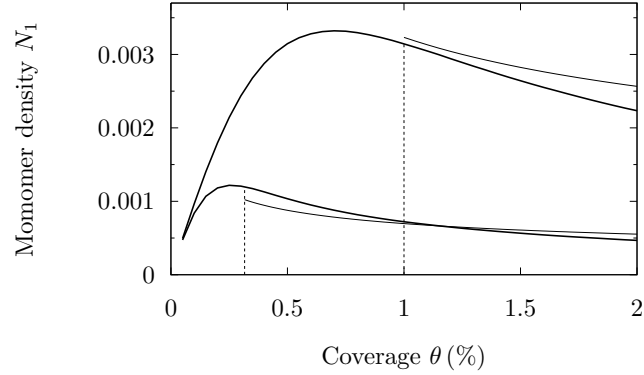}
\end{center}
\caption{Thick lines: the monomer densities obtained
in KMC simulations for two values of the diffusion to deposition rates
ratio $R=10^4$ (upper line) and $R=10^5$ (lower line); thin solid lines:
the densities in the steady-state regime calculated according to the
asymptotic formula  (\ref{ sgm1_1}); vertical dashed lines show the
positions of the crossover coverage $\theta^*$ from  (\ref{ crossover}).}
\label{fig2}
\end{figure}

To illustrate the above formalism, in this section we present the calculation
of $\sigma_1$ with the use of the KMC simulations and the formula
\begin{equation}
\label{ sgm1}
\sigma_1 = \sum_{i_{\rm NN}\in{\rm NN}}m_{i_{\rm NN}}
\langle\hat{N}_{1,i_{\rm NN}}\hat{N}_{1,i}/(MN_1^2)
\end{equation}
that can be derived from the last term in (\ref{ monomers_t}), as
explained in \ref{REs}.  As was noted in the Introduction, $\sigma_1$ is
the most important of the CNs in the systems belonging to the so-called
i=1 case when all islands with $s>1$ are stable.  This is because in
the usually considered case of large $R$ when the direct impingement
terms are small, $\sigma_1$ effectively defines the nucleation
of dimers and hence the total number of islands in the system because
every island nucleates as a dimer in the $i=1$ case \cite{bales_chrzan}.
The total island density measured experimentally can be used to define
the diffusivity of atomic monomers on the surface \cite{brune_et_al}.
Therefore, an accurate theoretical calculation of $\sigma_1$ could
lead to more reliable empirical estimates of the microscopic parameters
that define the surface diffusivity.  Because of this, $\sigma_1$ was
calculated by several authors but controversies remained.  For example,
the widely cited analytic expression \cite{tang93,1992,bales_chrzan,2006}
\begin{equation}
\label{ sgm1_0}
\sigma_1 \simeq \frac{4\pi}{\ln(CRN_1)},
\end{equation}
where $C$ is a numerical constant.  In the steady-state
regime at large $R$ the monomer density behaves as \cite{1992}
\begin{equation}
\label{ N_1}
N_1\sim (3\pi^2\theta R^2)^{-1/3}.
\end{equation}
It should be noted that this formula is non-physical at small coverage
where $N_1\approx\theta$ while according to  (\ref{ N_1}) $N_1$
diverges.  Therefore, we will use it only above the crossover
coverage separating the transient and the steady-state regimes \cite{2006}
\begin{equation}
\label{ crossover}
\theta^*=R^{-1/2}
\end{equation}
(see \fref{fig2}).  Thus, substituting  (\ref{ N_1}) into (\ref{ sgm1_0})
one gets
\begin{equation}
\label{ sgm1_1}
\sigma_1 \simeq \frac{4\pi}{\ln C_1+(\ln R-\ln\theta)/3},
\end{equation}
where $C_1=C/(3\pi^2)^{1/3}$.  As is seen, the dependence on $R$
in  (\ref{ sgm1_1}) is formally as strong as on $\theta$ and in
practice is even stronger because in experiments and simulations on
growth in precoalescence regime the coverage usually varies within one
order of magnitude ($\theta\sim1-20\%$) while $R$ varies over several
orders of magnitude. For example, in simulations of
\cite{korner_island_2010}) $R=10^5-10^8$.  Yet the authors arrive at the
conclusion that the coverage dependence is more important.

\begin{figure}
\begin{center}
\includegraphics[viewport = 110 0 300 380, scale=0.55]{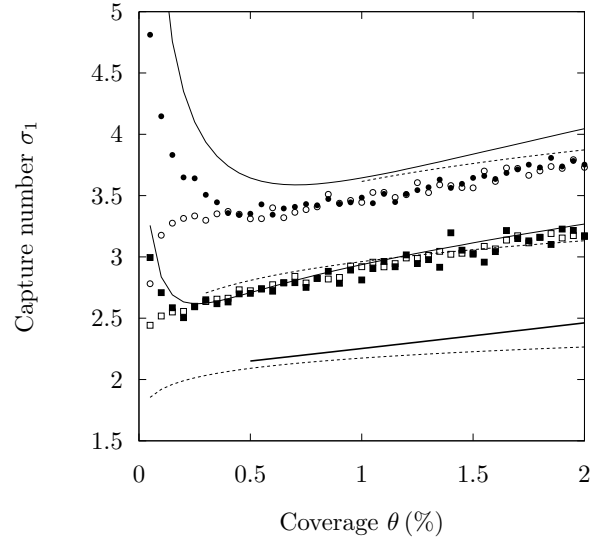}
\end{center}
\caption{\label{fig3}Filled symbols: KMC measurements with the use
of the exact formula (\ref{ exact}) at diffusion to deposition rates
ratios $R=10^4$ ($\bullet$) and $R=10^5$ ($\blacksquare$); similar
empty symbols: KMC measurements with the use of the procedure described
in \cite{1996} and \cite{korner_island_2010}. Thin
solid lines: $\sigma_1$ calculated according to  (\ref{ sgm1_0})
with $C=1$ \cite{bales_chrzan} with the use of the simulated values of
$N_1$ (see \fref{fig2}); dashed lines:  (\ref{ sgm1_1}) for
$\theta>\theta^*$. Thick solid line: fit to the $R=10^7$ KMC data
from figure 2 of \cite{korner_island_2010}.}
\end{figure}

To compare the values of $\sigma_1$ calculated in the framework
of the above theories with the exact values given by  (\ref{ sgm1})
we performed KMC simulations at two values of $R$ and at small coverage
shown in figures \ref{fig2} and \ref{fig3}.  The choice of parameters was
dictated by the necessity to obtain reasonable statistics which depend on the monomer density $N_1$, as can be seen from  (\ref{ sgm1}).  The latter diminishes with the growth of both $R$ and $\theta$, as is seen from  (\ref{ N_1}).  Therefore, we have chosen the region near the crossover coverage  (\ref{ crossover}) where the density is at its maximum (see \fref{fig2}) \cite{2006}.

In the harder $R=10^5$ case, 400 KMC runs in the system with
$4096\times4096$ sites were made.  In the easier $R=10^4$ case
$2048\times2048$ systems were used and lesser statistics were gathered to
obtain similar statistical errors.  No finite size effects were noticed.
As can be seen from \fref{fig3},  (\ref{ sgm1_0}) with the value $C=1$
taken from \cite{bales_chrzan} semiquantitatively describes the simulated
data over the whole range of coverage studied both in the transient regime
and in the course of the steady-state growth, especially at the larger
diffusion to deposition rates ratio $R$, as predicted theoretically. In
the paper \cite{tang93} similar value of $C\approx1.38$ was found.  Thus,
the results of  \cite{tang93,bales_chrzan} and our results seems to rule
out the values $C\sim32-64$ advocated in \cite{li_evans05,2006}.

On the other hand,  the KMC procedure suggested in \cite{1996} and
\cite{korner_island_2010} produces results in qualitative disagreement
with the exact data in the transient region, though practically coincides
with them within the statistical errors in the steady-state regime,as
can be seen from \fref{fig3}.  The reason for this seems to be clear.
The procedure in question consists in stopping the deposition from the
external source at any given moment and detaching the monomers captured by
the islands or separating the nucleated dimer and putting them back into
the homogeneous deposition flux over the free sites without occupied
neighbours \cite{1996,korner_island_2010}.  The capture/nucleation
events are counted for every island size. By continuing this arrested
growth simulation sufficiently long one can obtain arbitrarily good
statistics on capture numbers.  Obviously, however, that this procedure
may give quite accurate results only in the steady state regime where
the diffusion profile $N_1\mathbf{(x)}$ is time-independent, so the
origin of the homogeneously deposited atoms,---either from the physical
deposition flux or from the artificially simulated one,--- is irrelevant:
the profile is always the same \cite{1Dscaling}.

At small coverages at the early stage of the growth, however, the
diffusion profile only strives to reach the steady-state shape and so
is time dependent.  Thus, at the start of the arrested deposition the
diffusion profile has a physical shape which permanently changes with
time and as the arrested deposition simulation continues the shape of
the profile in the absence of the external deposition became unphysical
and the simulation produces incorrect $\sigma_1$ values, as can be seen
from \fref{fig3}.
\section{\label{conclusion}Conclusion}
In this paper we used the second quantization representation of the
stochastic master equation governing the coherent epitaxial growth
to develop a formalism that allows for a rigorous derivation of
the rate equations for the epitaxial growth.  The REs obtained
in sec.\ \ref{evolution} are more detailed than those found
in literature in that they describe the evolution of individual
island configurations.  This may be of value in the cases when the
number of configurations is restricted, as in the case of 1D growth
\cite{nat_chains,nat_chains2,albao_kinetic_2009}.  In the majority of
cases, however, the number of the island configurations is too large
to be used in practical calculations \cite{jensen_enumerations_2001}.
Therefore, it is conventional to use the REs where the island densities
comprising all island configurations of the same size are used.
We illustrated the derivation of such REs in section \ref{REs}.  In some
cases, however, experimental data show that not all configurations
are equal.  The so-called magic clusters may play more significant role
than islands of other shapes \cite{magicSi98,magicGeSi}. Our equations
of Sec.\ \ref{evolution} are capable of assuring a special treatment to
such configurations. Furthermore, in the case of quantum dot growth,
islands of different morphologies exhibit qualitatively different
growth behaviours \cite{rmp_dots,JPhys13} which also would require more
detailed characterization of islands than only by their size.  Finally,
the detailed derivation of the REs can be helpful in better understanding
the approximations underlying the REs and the means of improving them
with the use of more accurate decouplings.

As we showed in the present paper, even in the simplest case of the
irreversible submonolayer growth in precoalescence regime that has
been extensively investigated for about forty years there still remain
controversial issues that need further investigation.  Obviously that in
more complex cases such as the currently hot subject of the quantum dot
growth \cite{rmp_dots,JPhys13} the difficulties will enlarge manifold.
We expect that the formalism developed in the present paper will prove
to be an indispensable tool of derivation of the REs for such cases.  The
derivation in the framework of the second quantisation formalism, though
cumbersome, is quite straightforward. Besides, it is very flexible and
do not reduces to the simple case of the hard core bosons we considered
in the present paper.  Multiple site occupation, coalescence, 3D growth,
reversibility, {\em etc} all can be accounted for both in the quantum
formalism  \cite{doi,siggia,grassberger_scheunert,mattis_glasser,PRE96}
and in the REs \cite{Dobbs_ea97,reversible}.
\appendix
\section*{\label{commutators}Appendix}\setcounter{section}{1}
The commutators in  (\ref{ exact}) can be calculated with the use of the
chain rule
\begin{eqnarray}
\left [ A_1A_2\dots A_k,X \right ]=&&\left [ A_1,X \right ]A_2\dots A_k
+A_1\left [ A_2,X \right ]A_3\dots A_k+\dots\nonumber\\
&&+A_1\dots A_{k-1}\left [ A_k,X  \right ],
\label{ chain}
\end{eqnarray}
the definitions
\begin{equation}
\label{ nnbar}
n_i=a^+_ia_i,\quad \bar{n}_i\equiv 1-n_i=a_ia^+_i
\end{equation}
and the commutators
\begin{eqnarray}
\label{ na+}
&&\left [ n_i, a^+_i\right ]=-\left [ \bar{n}_i, a^+_i\right] =a^+_i\\
\label{ na}
&&\left [ n_i, a_i\right ]=-\left [ \bar{n}_i, a_i\right] =-a_i.
\end{eqnarray}

As can be easily seen, in the case of binary site occupancy
vector $|\Omega\rangle$ in  (\ref{ Omega})
can be expressed through vectors (\ref{ binary}) as
\begin{equation}
\label{ Omega2}
|\Omega\rangle=\prod_i(|0\rangle_i+|1\rangle_i).
\end{equation}
From this representation it is easy to see that inside the matrix elements
of type (\ref{ Omean2}), i.\ e., with vector $|\Omega\rangle$ on the left,
the nondiagonal operators can always be turned into diagonal ones as
\begin{eqnarray}
\label{ weaka+}
&&\langle\Omega|a^+_i=\langle\Omega|\bar{n}_i\\
\label{ weaka}
&&\langle\Omega|a_i=\langle\Omega|n_i.
\end{eqnarray}
Care should be taken to keep noncommuting operators in correct order.
For example, vital to the results below is that operator $\tilde{a}_i$
in  (\ref{ T}) is the rightmost in the product.  The evolution equations
that we derive strongly depend on this because of the following not
quite intuitive identities:
\begin{eqnarray}
\label{ attention}
&&\langle\Omega|\bar{n}_ia_i=\langle\Omega|n_i\\
&&\langle\Omega|\bar{n}_ia^+_i=0
\end{eqnarray}
[cf. (\ref{ weaka+})---(\ref{ weaka})].

Because physically there exists a qualitative difference between the islands
and mobile monomers, it should be reflected in the formalism.  Therefore, we
will discuss the calculation of the commutators for these entities separately.
\subsection{The island commutators ($s>1$)}
First we calculate the simpler commutator of the deposition term in
(\ref{ totalW}) with the island operator  (\ref{ islandN}).  From  (\ref{
na+}) it is seen that the commutator of the creation operators with
the island operator  (\ref{ islandN}) will consist of two parts :
the commutator with operators $n_i$ on interior island sites and with
operators $\bar{n}_i$ on the border sites. Though the commutators in
(\ref{ na+}) differ only by sign, the results are somewhat different.
The border commutators when acting on the vector $\langle\Omega|$
result in the same expression as the island operator  (\ref{ islandN})
itself while the interior commutators lead to islands with one less atom
that arise when every $n_i$ is sequentially replaced with $\bar{n}_i$
corresponding to an empty site.  We denote symbolically the set of such
islands as $\{c\setminus1\}$:
\begin{equation}
\label{ Fcommutator}
\langle\Omega|[\hat{N}^{(c)}_{s,i},\mathbf{W}^F]=
F\langle\Omega|\left(\sum_{\{c\setminus1\}}\hat{N}^{(c\setminus1)}_{s-1,i}
-\bar{s}\hat{N}^{(c)}_{s,i}\right),
\end{equation}
where the first sum is over $s$ islands whose configurations differ from
$c$ by the absence of each of $s$ atoms constituting the initial island.
It is to be noted that in some cases this might lead to disconnected sets,
i.\ e., to the appearance of two or more separate islands or monomers.
Such cases correspond to the growth with the coalescence and we do
not consider them in the present paper devoted to pre-coalescence
growth regime.  Their inclusion would lead to exact equations of more
general type.

The calculation of the commutator with the second term in  (\ref{
totalW}) in the case of islands is very similar due to the structure of
$\tilde{a}_i$ in  (\ref{ W_D}).  Because the annihilation operator in
$\mathbf{W}^D$ is multiplied by factors $\bar{n}_{i_{\rm NN}}$ on all
neighbour sites, $\tilde{a}_i$ will commute with any diagonal operator
on site $i$ that contains at least one of $n_{i_{\rm NN}}$ as a factor.
But such are all operators entering $\hat{N}^{(c)}_{s,i}$ by definition.
Therefore, in the commutator $\tilde{a}_i$ can be treated as a number
similar to $F$ and the commutator calculated exactly as in  (\ref{
Fcommutator}).  The main complication is connected with acting with
D$\sum_{i_{\rm N}}\tilde{a}_{i_{\rm N}}$ on the vector $\langle\Omega|$.
But as explained above, for every boundary site only the sites that have
no neighbours with $n_i$ will contribute, that is, only the NN sites to
the island with configuration $c$ and the NN sites to the islands with
one less atom [see  (\ref{ Fcommutator})] amounting to
\begin{equation}
\label{ Dcommutator}
\fl\langle\Omega|[\hat{N}^{(c)}_{s,i},\mathbf{W}^D]=D
\sum_{\{c\setminus1\}i_{\rm NN}}^{\qquad\prime}\langle\Omega|
\hat{N}_{1,{i_{\rm NN}}}\hat{N}^{(c\setminus1)}_{s-1,i}
-D\langle\Omega|\sum_{i_{\rm NN}}m_{i_{\rm NN}}\hat{N}_{1,{i_{\rm NN}}}\hat{N}^{(c)}_{s,i},
\end{equation}
where the prime over the first sum is to indicate that the summation
over the NN sites comprises only those sites (there may be from one to
three) which the detached atom can rich in one hop and $m_{i_{\rm NN}}$
is the number of nucleation paths for the NN atom to reach the
island \cite{reversible}.
\subsection{The monomer case}
Because the monomer is a limiting case of an island, we can proceed
with the calculation of commutator as above by only bearing in mind
that monomers differ from islands in two respects.  First, when an
atom is taken away from the monomer in the first terms in (\ref{
Fcommutator}) and (\ref{ Dcommutator}) we obtain an ``empty island".
In the first case we find
\begin{equation}
\label{ Fcommutator1}
\langle\Omega|[\hat{N}_{1,i},\mathbf{W}^F]=
F\langle\Omega|\left(\hat{N}_{0,i} -4\hat{N}_{1,i}\right),
\end{equation}
where the ``empty island" operator 
\begin{equation} \label{ empty}
\hat{N}_{0,i}=\bar{n}_i\prod_{\hat{e}=\pm\hat{x},\pm\hat{y}}\bar{n}_{i+\hat{e}}
\end{equation} 
has formal structure of $\hat{N}_{1,i}$ \eref{ N_1i} shown in \fref{fig1}
but with the central operator $n_i$ replaced with $\bar{n}_i$.
The second commutator takes the form
\begin{equation}
\label{ Dcommutator1}
\fl\langle\Omega|[\hat{N}_{1,i},\mathbf{W}^D]=
\langle\Omega|(-4D\hat{N}_{1,i})
+\langle\Omega|D\left(\sum_{i_{\rm N}}
\hat{N}_{1,i_{\rm N}}\hat{N}_{0,i}^{({i_{\rm N}})}
-\sum_{i_{\rm NN}}m_{i_{\rm NN}}\hat{N}_{1,i_{\rm NN}}\hat{N}_{1,i}
\right),
\end{equation}
where the last two terms in parentheses are analogous to the terms
entering (\ref{ Dcommutator}) for the islands while the first term is
due to the commutator with $\tilde{a}$ from $W^D$  (\ref{ W_D}) that
is absent in the island case.  The operator $\hat{N}_{0,i}^{({i_{\rm
N}})}$ is similar to the empty island operator $\hat{N}_{0,i}$  and
can be obtained from it by means of  (\ref{ attention}), namely by
replacing one of $\bar{n}_{i_{\rm N}}$ factors surrounding site $i$
with $n_{i_{\rm N}}$.  Its action is to suppress those configurations
in which an atom detaches from an island.
\ack
One of the authors (V. I. T.) expresses his gratitude to Universit{\'e}
de Strasbourg and IPCMS for their hospitality.

\end{document}